\documentclass[twocolumn,aps,prb]{revtex4}
\usepackage{graphicx}

\begin{document}
\author{Yaroslav Tserkovnyak and Arne Brataas} \affiliation{Harvard
University, Lyman Laboratory of Physics, Cambridge, Massachusetts 02138}
\title{Shot noise in ferromagnet--normal metal systems}

\begin{abstract}
A semiclassical theory of the low frequency shot noise in
ferromagnet--normal metal systems is formulated. Noncollinear magnetization
directions of the ferromagnetic leads, arbitrary junctions and the
elastic and inelastic scattering regimes are considered. The shot
noise is governed by a set of mesoscopic parameters that are expressed
in terms of the microscopic details of the junctions in the
circuit. Explicit results in the case of ballistic, tunnel, and
diffusive junctions are evaluated. The shot noise, the current and the
Fano factor are calculated for a double barrier ferromagnet--normal
metal--ferromagnet system. It is demonstrated that the shot noise can
have a nonmonotonic behavior as a function of the relative angle
between the magnetizations of the ferromagnetic reservoirs.
\end{abstract}

\maketitle

\section{Introduction}

Magnetoelectronic circuits have recently attracted considerable
interest due to their potential for magnetic random access memories,
sensors, and for fundamental studies of spin transport in magnetic and
nonmagnetic
devices.\cite{Meservey94:173,Levy94:367,Gijs97:285,Prinz98:282} A
ferromagnetic metal in contact with a paramagnet can inject spins into
the paramagnet or detect spins from the
paramagnet.\cite{Johnson85:1790} Spin accumulation and its effect on
the current-voltage characteristics have been studied thoroughly in
double barrier ferromagnet--normal metal--ferromagnet (\textit{F-N-F})
systems.\cite{Brataas00:2481, Barnas98:85}

An important parameter in view of applications of spin injection is
the noise to signal ratio. The noise furthermore provides additional
information on the electronic structure and the nonequilibrium
transport processes and that is our main motivation for this
study. The calculation and experimental detection of shot noise has
been an active subfield of mesoscopic physics during the last
decade.\cite{Blanter00:1} Shot noise in systems consisting of normal
metals, superconductors, semiconductors and combinations thereof
combined by tunnel barriers, diffusive barriers and ballistic barriers
have been studied.\cite{Blanter00:1} There have been much less
attention on the shot noise in hybrid systems involving ferromagnets.

Only a few studies have been carried out on the effects of the
fluctuations in the spin accumulation on the shot noise in ferromagnet
- normal metal systems. Bulka \textit{et al.} computed in Ref.\
\onlinecite{Bulka99:12246} the shot noise in \textit{F-N-F} double barrier
systems in the Coulomb blockade regime when the magnetizations in the
ferromagnetic reservoirs are collinear and found pronounced and
interesting effects due to the interplay of the spin and charge
fluctuations . Their results in the regime when the source-drain bias
is larger than the Coulomb charging energy can be understood in terms
of well-known results for normal-metal systems for two spin
directions.\cite{Blanter00:1} Nowak \textit{et al.}\cite{Nowak99:600} measured
electrical noise in ferromagnet--insulator--ferromagnet (\textit{F-I-F})
systems in the collinear configurations. They also obtained results
consistent with a generalization of the results for the normal metal
systems. Spin-dependent tunneling through \textit{F-I-F} systems in the case of
noncollinear magnetization configurations were experimentally studied
by Moodera \textit{et al.}\cite{Moodera96:4724} with results in good agreement
with the predictions for the angular dependence of magnetoresistance
by Slonczewski,\cite{Slonczewski89:6995} but no noise measurements
were performed. It is our purpose to provide a more general discussion
of the shot noise in ferromagnet--normal metal systems where Coulomb
charging effects are negligible.

The noise power $S(\omega)$ is defined as
\begin{equation}
S(\omega)=2\int_{-\infty}^{\infty}dte^{-i\omega
t}\left\langle\Delta I(t)\Delta I(0)\right\rangle \, ,
\label{sn}
\end{equation}
where $\Delta I(t)=I(t)-\left\langle I(t)\right\rangle$ denotes the
fluctuation of the current $I(t)$ from its average $\left\langle
I(t)\right\rangle$. The noise power has contributions from $1/f$
noise, thermal noise, and shot noise. In equilibrium, the thermal
noise is given by the Johnson-Nyquist formula in terms of the
conductance of the system. We will consider the nonequilibrium noise
contributions in the limit when the bias voltage is larger than the
thermal energy $eV\gg k_{\rm B}T$, so that the thermal
Johnson-Nyquist noise can be disregarded. The $1/f$ noise dominates at
very low frequencies or high bias.

Although our theoretical framework allows the calculation of the shot
noise in multiterminal systems, we will focus our attention on double
barrier systems. Double barrier systems have a characteristic time
scale corresponding to the dwell time $\tau_d$ of the electron on the
node. We will consider the low-frequency shot-noise contribution to
the noise power when $\omega\ll\tau_d^{-1}$ but the frequency $\omega$
is higher than the frequency at which the $1/f$ noise dominates. Thus
in this regime the low-frequency shot noise is the dominant term and
we will set $\omega=0$ in the following and disregard the $1/f$ noise.

The Landauer-B\"{u}ttiker (LB) theory of transport in phase-coherent
mesoscopic conductors gives simple expressions for the average
current\cite{Buttiker86:1761}
\begin{equation}
I=\frac{e^2}{h}V\sum_n T_n \, ,
\label{LB-1}
\end{equation}
and the low-frequency shot-noise power
\begin{equation}
S=\frac{2e^3}{h}V\sum_n T_n(1-T_n) \, ,
\label{LB-2}
\end{equation}
where $V$ is the applied voltage and $T_n$ (transmission
probabilities) are the eigenvalues of $tt^\dagger$ composed from the
transmission matrix $t$ of the junction at Fermi energy. It is assumed
that the scattering matrix is energy independent in the energy
interval determined by the chemical potentials of the voltage
probes. This is easily realized in metallic systems and/or at low
bias. Of special interest is the Fano factor which is defined as the
ratio between the shot-noise power and the average current:
$F=S/(2eI)$. When the transmission of the particles of charge $e$ is
Poissonian (\textit{i.e.}, random) the shot noise is $S=2eI$, and the Fano
factor is $F=1$.

\begin{figure*}
\includegraphics[scale=0.85]{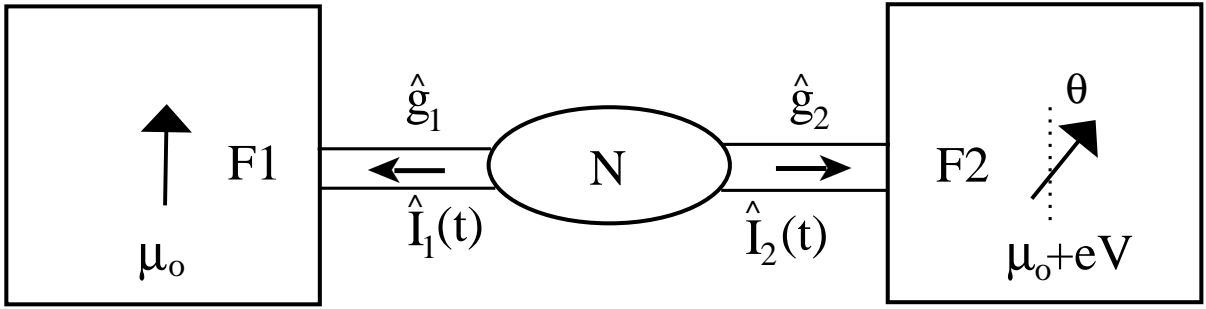}
\caption{The two-terminal device consisting of a normal metal node (\textit{N})
attached to two ferromagnetic reservoirs (\textit{F}1 and \textit{F}2)
with arbitrary relative magnetization direction $\theta$. A
source-drain bias $V$ is applied between the ferromagnetic reservoirs
and time-dependent currents $\hat{I}_1(t)$ and $\hat{I}_2(t)$
flow into the normal metal island from the reservoirs
\textit{F}1 and \textit{F}2, respectively.
The contact between the ferromagnetic \textit{F}1 (\textit{F}2)
and the normal metal node is characterized by the conductance
matrices $\hat{g}_1$ ($\hat{g}_2$).}
\label{dd}
\end{figure*}

We consider a double barrier \textit{F-N-F} system as shown in Fig.\
\ref{dd}. The system is composed of a chaotic cavity connected to two
voltage probes at different chemical potentials through junctions. The
left and the right probes are ferromagnetic with magnetization
directions ${\bf m}_1$ and ${\bf m}_2$, respectively. Our theory is
applicable if the following assumptions hold: The spin-flip relaxation
time on the normal metal island is much longer than the typical dwell
time on the island. The size of each of the two junctions is much
smaller than the corresponding phase-coherence length, but the motion
of the electrons inside the normal metal island can be elastic or
inelastic. The electrons trapped in the cavity dwell long enough so
that the scattering in the cavity leads to a homogeneous and isotropic
phase-space distribution of electrons inside the cavity. This can be
achieved, for example, by scattering off the bulk disorder (diffusive
cavity) or by scattering off the irregularities at the surface
(ballistic cavity). The theory is applicable if the cavity has
negligible resistance compared to the resistances of the junctions,
e.g. the voltage only drops across the junctions. Two scattering
regimes are considered: quasielastic when $\tau_d\ll\tau_{\text{in}}$, and
inelastic when $\tau_{\text{in}}\ll\tau_d$, where $\tau_{\text{in}}$ is the
inelastic scattering time, and $\tau_d$ is the dwell time in the
cavity.

We will study the influence of the relative difference in the
magnetization directions on the shot noise. In the case of collinear
magnetizations, the results for the all-normal metal systems can be
easily generalized by taking the two spin directions separately into
account. The regime of noncollinear magnetization configuration is
more interesting and we will present novel nontrivial results for
this case. There are, to the best of our knowledge, no systematic
predictions for the Fano factor in this regime. We will also present
results for the shot noise in other transport regimes than previously
studied for \textit{F-N} systems. In addition to the tunneling regime, already
discussed in some limits,\cite{Bulka99:12246,Nowak99:600} we will
extend the analysis to ballistic and diffusive junctions. Different
regimes of scattering on the normal metal node will also be studied:
The elastic scattering regime and the inelastic scattering regime.

Let us first review the results for transport through a single barrier
in all-normal metal systems.\cite{Blanter00:1} The Fano factor in the
cases of ballistic (\textit{B}), diffusive (\textit{D}), and tunnel (\textit{T}) junctions are
\begin{eqnarray}
F_B &= & 0 \, ,    \label{FanoB} \\
F_D &= & 1/3 \, , \label{FanoD} \\
F_T &= & 1   \, . \label{FanoT} 
\end{eqnarray}
In the ballistic and tunnel regimes the result follows directly from
the LB theory (\ref{LB-1},\ref{LB-2}), since $T_n\ll 1$ for tunnel
junctions and $T_n=0$ or $T_n=1$ for ballistic junctions. The $1/3$ noise
suppression in diffusive junctions is due to the bimodal form of the
distribution function of transmission eigenvalues.\cite{Blanter00:1}

Next, let us review the results for all-normal metal {\em double}
barrier systems in the elastic transport regime. The Fano factor is
$F=(R_1^3F_1+R_2^3F_2+R_1^2R_2+R_1R_2^2)/(R_1+R_2)^3$, where $F_i$
($R_i$) is the Fano factor (resistance) of the $i$th
junction.\cite{Beenakker92:1889} For ballistic, diffusive and tunnel
junctions the Fano factors for double barrier systems are
\begin{eqnarray}
F_{B,2}^{\text{el}} & = & \frac{R_1R_2}{(R_1+R_2)^2} \, ,
\label{F_EB2} \\
F_{D,2}^{\text{el}}& = &1/3 \, ,
\label{F_ED2} \\
F_{T,2}^{\text{el}}& = &\frac{R_1^2+R_2^2}{(R_1+R_2)^2} \, .
\label{F_ET2}
\end{eqnarray}
In the ballistic case (\ref{F_EB2}), each junction exhibits zero shot
noise, but the shot noise of the double barrier system is finite since
the chaotic nodes have a partial (fluctuating) occupation of the
energy levels in the nonequilibrium situation even at zero
temperature. Consequently, although the direct contribution to the
shot noise from a single junction vanishes, the effective 'thermal'
contribution to the noise is finite. The elastic chaotic cavity was
recently realized as a quantum dot connected to voltage probes with
two point contacts\cite{Oberholzer:01:2114} and the Fano factor
(\ref{F_EB2}) was experimentally confirmed. For diffusive junctions,
the Fano factor (\ref{F_ED2}) is the same as for a single junction
(\ref{FanoD}) because an elastic double diffusive barrier system can
be seen as a single diffusive junction. For tunnel junctions, the Fano
factor (\ref{F_ET2}) attains its minimum, $F_{T,2}^{\text{el}}=1/2$,
when the barriers are symmetric. For completely asymmetric barriers,
the system is governed by the barrier with the lowest transmission
only and therefore the Fano factor is $F_{T,2}^{\text{el}}=1$ as in
the case of a single tunnel barrier.

In the inelastic transport regime the Fano factor for a double barrier
system is\cite{Beenakker92:1889}
$F=(R_1^2F_1+R_2^2F_2)/(R_1+R_2)^2$. Because the energy level
occupation numbers in the node obey the Fermi-Dirac (FD) distribution
with a fluctuating chemical potential, the Fano factor of the entire
system is a linear combination of the Fano factors of the two
junctions. Ballistic, diffusive, and tunnel junctions then yield,
respectively,
\begin{eqnarray}
F_{B,2}^{\text{in}} &=& 0 \label{F_IB2} \, , \\ F_{D,2}^{\text{in}}
&=&
\frac{1}{3}\frac{R_1^2+R_2^2}{(R_1+R_2)^2}
\, , \label{F_ID2} \\ F_{T,2}^{\text{in}} &=&
\frac{R_1^2+R_2^2}{(R_1+R_2)^2} \, . \label{F_IT2}
\end{eqnarray}
In the case of tunnel junctions, we get the same result as in the
elastic regime. The Fano factor in the diffusive double barrier system
differs from $1/3$, which is a universal noise suppression factor in
elastic diffusive wires. $1/3$ suppression is restored only for a
completely asymmetric system governed by the lowest transmission
barrier. Finally, ballistic system has vanishing noise because the
node has a FD distribution of electrons.

We will generalize the results above to ferromagnet--normal metal
systems with noncollinear magnetization directions. The manuscript is
organized in the following way. In Sec.~\ref{s:circuit} we describe
the circuit theory used to analyze multiterminal \textit{F-N} systems and
relate the current and its fluctuations through a junction to the
adjacent distribution functions and the scattering matrices for
spin-up and spin-down electrons. We compute the Fano factor for double
barrier junction systems in Sec.~\ref{s:double}. Our conclusions
can be found in Sec.~\ref{s:con}.

\section{Circuit Theory}
\label{s:circuit}

Nazarov realized that transport in hybrid superconductor--normal
metal systems can be understood in terms of a generalized circuit
theory similar to Ohm's law.\cite{Nazarov94:134} In a similar way a
circuit theory for ferromagnet--normal metal systems was formulated
by Brataas, Nazarov, and Bauer.\cite{Brataas00:2481} We extend the
circuit theory of transport through ferromagnetic-normal metal systems
to the calculation of shot noise. A (magneto)electronic circuit is
divided into junctions (resistive elements), nodes (low impedance
interconnectors), and reservoirs (voltage probes at thermal
equilibrium). The theory is applicable when the junctions limit the
electric current and the nodes are characterized by a distribution
function which is constant in position space (this assumption can be
relaxed, see Ref.\ \onlinecite{Huertas00:5700}) and isotropic in
momentum space. It is assumed that there are no spin-flip processes in
the nodes. In the following discussion the nodes are taken to be
normal metal chaotic cavities. Each voltage probe is ferromagnetic,
connected to the nodes through leads carrying current and
spin current. The current in the circuit depends on the relative
magnetization orientations in the probes.\cite{Brataas00:2481}

The current in the circuit can be derived by using the Keldysh Green
function technique.\cite{Brataas00:2481} We obtain the same result
using LB approach generalized to describe circuits with one or more
terminals (the nodes) with nonequilibrium spin distribution
functions. The assumption of the LB theory is that particles exiting a
probe have the same occupation of the energy levels as the particles
inside the probe and the occupation numbers of the incoming particles
are uniquely determined by the probes that supply them and the
scattering properties of the junctions. We generalize the LB
expression for the current operator from a normal metal node into a
junction to a $2\times2$ current operator in spin space in order to
also describe spin currents collinear and noncollinear to the
magnetization directions in the adjacent ferromagnets.  The $2\times2$
current operator for the particle flow from node $2$ to node $1$ is
\begin{eqnarray}
\hat{I}^{\alpha\beta}(t)&=&\frac{e}{h}\sum_{n}\int dEdE^{\prime}e^{i(E-E^\prime)t/\hbar}\nonumber\\
&&\left[a_{\beta n,1}^\dagger(E)a_{\alpha n,1}(E^\prime)-b_{\beta n,1}^\dagger(E)b_{\alpha n,1}(E^\prime)\right] \, ,\nonumber\\
\label{I}
\end{eqnarray}
where $a_{\alpha n,i}^\dagger(E)$ creates a spin-$\alpha$ electron
with energy $E$ leaving the $i$th node through the
$n$th transverse mode and $b_{\alpha n,i}^\dagger(E)$
creates a spin-$\alpha$ electron with energy $E$ entering the
$i$th node through the $n$th transverse mode. The
current operator is $I=\sum_{\alpha} \hat{I}^{\alpha\alpha}$ and the
spin-current operator is $\hat{\bf I}_s = -\hbar/(2e)\sum_{\alpha\beta}
\hat{\text{\boldmath $\sigma$}}_{\alpha \beta} \hat{I}^{\beta\alpha}$,
where $\hat{\text{\boldmath
$\sigma$}}=(\hat{\sigma}_x,\hat{\sigma}_y,\hat{\sigma}_z)$ is a vector
of the Pauli matrices. If the scattering matrices of the junctions are
known then we can express annihilation operators $b_{\alpha n,i}(E)$
in terms of the operators $a_{\alpha n,i}(E)$. Expectation values
involving operators $a_{\alpha n,i}^\dagger(E)$ and $a_{\alpha
n,i}(E)$ can then be evaluated in terms of the distribution functions
in the adjacent nodes and we can calculate both the current and the
shot noise.\cite{Buttiker86:1761}

Let us consider a node that is connected to several other nodes via
junctions. In the low frequency, long-time limit $\omega \ll 1/\tau_d$
the charge and spin in the node are conserved at every instance of
time
\begin{equation}
\sum_i \hat{I}_i(t)=0 \, ,
\label{si}
\end{equation}
where $\hat{I}_i$ is the current flowing out of the node through the
$i$th junction at time $t$ and the index $i$ runs over all
junctions connected to the node. Due to the charge and spin
conservation (\ref{si}), the distribution function on the normal metal
node has two contributions: A stationary part $\hat{f}_N(E)$ and a
small fluctuating part $\delta\hat{f}_N(E,t)$ caused by the
discreteness of the charges and spins that enter and leave the
island. Using the isotropy assumption in momentum space, the
stationary part $\hat{f}_N(E)$ is defined by $\left\langle a_{\alpha
m}^{\dagger }(E)a_{\beta
n}(E^\prime)\right\rangle=\delta_{nm}\delta(E-E^\prime)
\hat{f}_N^{\beta\alpha}(E)$, where $a_{\alpha m}^{\dagger}$ creates a
spin-$\alpha$ particle in the $m$th quantum state of the
node. The distribution function $\hat{f}_N(E)$ is a $2\times2$
Hermitian matrix in spin space to allow a nonequilibrium
spin accumulation on the normal metal node.

Following Beenakker and
B\"{u}ttiker,\cite{Beenakker92:1889,Blanter00:1} we separate the
time-dependent current fluctuations $\Delta\hat{I}_i(t)$ in the
$i$th junction into two contributions. The first contribution
$\delta\hat{I}_i(t)$ is due to the intrinsic noise in the junction if
the fluctuating term $\delta\hat{f}_N(E,t)$ of the distribution
function on the node is neglected. This is the only relevant term for
the shot noise for a single junction connected to reservoirs that have
stationary distribution functions. The second contribution is due to
the small fluctuation in the occupation of the energy levels on the
node $\delta\hat{f}_N(E,t)$.

Spin and charge conservation (\ref{si}) dictate the average occupation
of the energy levels in the node in terms of the average distribution
functions in the adjacent nodes.  From Eq.\ (\ref{si}) the total
fluctuations of the current vanish
\begin{equation}
\sum_i\Delta\hat{I}_i(t)=0 \, . 
\label{cs}
\end{equation}
Consequently, small fluctuations of the occupation of the energy
levels are needed to compensate for the intrinsic current fluctuations
in the junctions.\cite{Beenakker92:1889,Blanter00:1} The total
fluctuation $\Delta\hat{I}_i(t)$ is a sum of these two contributions:
\begin{equation}
\Delta\hat{I}_i(t)=\delta\hat{I}_i(t)+\sum_j\frac{\delta\langle\hat{I}_i\rangle}{\delta\langle\hat{f}_j\rangle}\delta\hat{f}_j(t) \, ,
\label{dI}
\end{equation}
where $j$ runs over the nodes adjacent to the $i$th
junction. The calculation of the term proportional to
$\delta\hat{f}_j(t)$ resembles the case of all-normal metal systems,
but is now generalized to describe the $2\times2$ spin-space
matrices. The calculation of this term is straightforward, but tedious
since it requires solving for the $2 \times 2$ distribution functions
in the circuit. The explicit result for the total fluctuation in the
current $\Delta \hat{I}_i(t)$ in terms of the intrinsic fluctuation in
the current through the separate junctions $\delta \hat{I}_i(t) $ is
shown below in Eq.\ (\ref{di}) for symmetric double barrier systems.

The ferromagnetic reservoirs are in thermal equilibrium characterized
by equilibrium FD distribution functions independent of time. In
contrast, the normal metal islands are isolated from the thermal baths
and have (fluctuating) nonequilibrium distribution functions required
for current conservation. The normal metal nodes can be viewed as
fictitious probes.\cite{Beenakker92:1889} They are characterized by an
isotropic and homogeneous distribution function like the true probes
(\textit{i.e.}, probes in local thermal equilibrium), but the (fluctuating)
distribution function does not have to be of the FD form.

\subsection{Average current}

Naturally, we obtain here the same result for the average current by
using the LB formalism as from the Keldysh
formalism\cite{Brataas00:2481} since they are identical when there is
no inelastic scattering in the junctions. For completeness we show the
result for the average current (per unit of energy, at a given energy)
between a ferromagnetic node and a normal metal node at the normal
metal side:\cite{Brataas00:2481}
\begin{eqnarray}
\hat{\imath}&=&\frac{e}{h}\left[g^{\uparrow}\hat{u}^{\uparrow}(\hat{f}_F-\hat{f}_N)\hat{u}^{\uparrow}+g^{\downarrow}\hat{u}^{\downarrow}(\hat{f}_F-\hat{f}_N)\hat{u}^{\downarrow}\nonumber\right.\\
&&\hspace{0.5cm}\left.-g^{\uparrow\downarrow}\hat{u}^{\uparrow}\hat{f}_N\hat{u}^\downarrow-g^{\downarrow\uparrow}\hat{u}^{\downarrow}\hat{f}_N\hat{u}^{\uparrow}\right] \, ,
\label{c}
\end{eqnarray}
where $\hat{u}^{\uparrow}=(\hat{1}+\hat{\text{\boldmath
$\sigma$}}\cdot{\bf m})/2$ and $\hat{u}^{\downarrow
}=(\hat{1}-\hat{\text{\boldmath $\sigma$}}\cdot{\bf m})/2$ are
projection matrices, $\hat{f}_N$ is the average distribution function
in the normal metal node, $\hat{f}_F$ is the equilibrium distribution
function in the ferromagnetic probe, and the conductances are defined
in terms of the reflection matrices for electrons incoming from the
normal metal node
\begin{eqnarray}
\hat{g}&=&\left(
\begin{array}{cc}
g^{\uparrow } & g^{\uparrow \downarrow } \\ 
g^{\downarrow \uparrow } & g^{\downarrow }
\end{array}
\right)\nonumber\\ &=&
\sum_{nm}\left( 
\begin{array}{cc}
\delta_{nm} - |r_{nm}^{\uparrow }| ^{2} & \delta_{nm} -
r_{nm}^{\uparrow} (r_{nm}^{\downarrow})^{\ast} \\ \delta_{nm} -
r_{nm}^{\downarrow} (r_{nm}^{\uparrow})^{\ast} & \delta_{nm} -
|r_{nm}^{\downarrow }| ^{2}
\end{array}
\right) \, . \nonumber\\
\label{g}
\end{eqnarray}
For energies much less than the Fermi energy, the conductance
(\ref{g}) is energy independent. Therefore, the elastic and inelastic
transport regimes are equivalent for the average current. However, the
shot noise differs in these transport regimes, since it is sensitive
to the the energy-resolved distribution.\cite{Blanter00:1}

\subsection{Shot noise}

The starting point in calculating the noise is the relation between
the fluctuation of the current in a junction and of the distribution
function in the fictitious probes adjacent to the junction
(\ref{dI}). The first term on the right-hand side of the Eq.\
(\ref{dI}) is a spontaneous Langevin source. The second term
(\ref{dI}) can be found in terms of the Langevin source by using
Eq.\ (\ref{cs}) for the instantaneous conservation of spin and
Eq.\ (\ref{c}) for the current. The problem of calculating the noise is
then reduced to finding the correlator of the Langevin sources
$\delta\hat{I}_i$ in terms of the scattering matrices of the junctions
and the distribution functions in the adjacent nodes and reservoirs.

The correlators of the Langevin sources can be found by generalizing
the LB theory for all-normal metal systems. We expand the current in
terms of the creation and annihilation operators, using Eq.\
(\ref{I}), and then determine particle and spin fluctuations
\begin{eqnarray}
&&\left\langle a_{\alpha k}^\dagger(E_{1})a_{\beta l}(E_{2})a_{\gamma
m}^\dagger(E_{3})a_{\delta
n}(E_{4})\right\rangle\nonumber\\
&&\hspace{0.5cm}-\left\langle a_{\alpha
k}^\dagger(E_{1})a_{\beta l}(E_{2})\right\rangle\left\langle a_{\gamma
m}^\dagger(E_{3})a_{\delta n}(E_{4})\right\rangle\nonumber\nonumber\\
&&\hspace{0.5cm}=\delta_{kn}\delta_{lm}\delta(E_{1}-E_{4})\delta(E_{2}-E_{3})\nonumber\\
&&\hspace{0.5cm}\times\hat{f}^{\delta\alpha}_n(E_{1})\left[1-\hat{f}^{\beta\gamma}_m(E_{2})\right] \, ,
\label{cccc}
\end{eqnarray}
where the Greek subscripts denote spin indices, and the Latin
subscripts include both the transverse mode index and the probe
label. However, the distribution function in the probe $\hat{f}_n$ is
independent of the transverse mode and $n$ here denotes only the probe
label.

A similar analysis was used to calculate shot noise in nonmagnetic
circuits.\cite{Beenakker92:1889} Our treatment of spin transport
is a generalization to include the spin degree of freedom. In addition
to instantaneous conservation of charge in the nodes, we use
instantaneous conservation of spin. Both conditions are satisfied when
the frequency is lower than the inverse dwell time of a particle in
the node, $\omega\tau_d\ll 1$.

When the inelastic scattering time is much longer than the transport
dwell time, the energy of a quasiparticle is conserved. In this
regime both current and noise can be calculated at each energy level
and then integrated over the energy to get the final result.

The inelastic regime is achieved when the inelastic scattering time is
shorter than the transport dwell time. When a particle is trapped in a
node, it is assumed to drop to the lowest energy state allowed without
a spin flip. As a result, there will be a direction of
spin accumulation in the node with FD distribution for electrons
polarized parallel and antiparallel to it (with different chemical
potentials $\mu^\uparrow$ and $\mu^\downarrow$, respectively). The
direction of spin accumulation and its magnitude are determined by the
transport rates integrated over all energies.

\subsection{Junction conductance and shot-noise matrices}

Because the current (\ref{I}) is expressed as a linear combination of
products of two creation and annihilation operators, the current in
the circuit is completely determined by $2\times 2$ conductance
matrices (\ref{g}), elements of which can be expressed as traces of
two reflection matrices.
Shot noise is quadratic in current, and we expect that it will be
governed by similar traces with four reflection and transmission
matrices. We show that this is indeed true with some
simplifications. First, because ferromagnetic probes are in local
thermal equilibrium, the new independent parameters can be expressed
in terms of the traces of four reflection matrices only. Second, only
the traces of the form
$\mbox{Tr}[r^{\alpha}(r^{\alpha^\prime})^\dagger
r^{\beta}(r^{\beta^\prime})^\dagger]$
enter the shot noise, where $\alpha$, $\alpha^\prime$, $\beta$, and
$\beta^\prime$ denote the spin. The new set of parameters describing
the shot noise of a junction can therefore be grouped into a $4\times
4$ Hermitian matrix $\hat{s}$:
\begin{widetext}
\begin{equation}
\hat{s}=\mbox{Tr}\left[\left(
\begin{array}{cccc}
\hat{1} & \hat{1} & \hat{1} & \hat{1}\\ \hat{1} & \hat{1} & \hat{1} &
\hat{1}\\
\hat{1} & \hat{1} & \hat{1} & \hat{1}\\ \hat{1} & \hat{1} & \hat{1} &
\hat{1}
\end{array}
\right)-\left(
\begin{array}{cccc}
r^{\uparrow}
(r^{\uparrow})^{\dagger}r^{\uparrow}(r^{\uparrow})^{\dagger} &
r^{\uparrow}
(r^{\uparrow})^{\dagger}r^{\uparrow}(r^{\downarrow})^{\dagger} &
r^{\uparrow}
(r^{\downarrow})^{\dagger}r^{\uparrow}(r^{\uparrow})^{\dagger} &
r^{\uparrow}
(r^{\downarrow})^{\dagger}r^{\uparrow}(r^{\downarrow})^{\dagger}\\

r^{\uparrow} (r^{\uparrow})^{\dagger}r^{\downarrow}
(r^{\uparrow})^{\dagger} & r^{\uparrow}
(r^{\uparrow})^{\dagger}r^{\downarrow} (r^{\downarrow})^{\dagger} &
r^{\uparrow} (r^{\downarrow})^{\dagger}r^{\downarrow}
(r^{\uparrow})^{\dagger} & r^{\uparrow}
(r^{\downarrow})^{\dagger}r^{\downarrow} (r^{\downarrow})^{\dagger}\\

r^{\downarrow} (r^{\uparrow})^{\dagger}r^{\uparrow}
(r^{\uparrow})^{\dagger} & r^{\downarrow}
(r^{\uparrow})^{\dagger}r^{\uparrow} (r^{\downarrow})^{\dagger} &
r^{\downarrow} (r^{\downarrow})^{\dagger}r^{\uparrow}
(r^{\uparrow})^{\dagger} & r^{\downarrow}
(r^{\downarrow})^{\dagger}r^{\uparrow} (r^{\downarrow})^{\dagger}\\

r^{\downarrow} (r^{\uparrow})^{\dagger}r^{\downarrow}
(r^{\uparrow})^{\dagger} & r^{\downarrow}
(r^{\uparrow})^{\dagger}r^{\downarrow} (r^{\downarrow})^{\dagger} &
r^{\downarrow} (r^{\downarrow})^{\dagger}r^{\downarrow}
(r^{\uparrow})^{\dagger} & r^{\downarrow}
(r^{\downarrow})^{\dagger}r^{\downarrow} (r^{\downarrow})^{\dagger}
\end{array}
\right)\right] \, ,
\label{spara}
\end{equation}
\end{widetext}
where the trace is taken for each element of the $4\times 4$ matrix
(each of which is an $M\times M$ matrix in the basis of the $M$
transverse modes) inside the square brackets, \textit{i.e.}, the result of the
operation is a $4\times 4$ matrix of complex numbers.

Only four off-diagonal elements and three diagonal elements are
independent in the shot-noise matrix $\hat{s}$. Therefore, in general,
we introduce 7 new parameters: The real parameters $s_\uparrow$,
$s_{\uparrow\downarrow}$, $s_\downarrow$, and the complex quantities
$s_+$, $s_-$, $s_0$, $\bar{s}_0$,
\begin{equation}
\hat{s}=\left(
\begin{array}{cccc}
s_\uparrow & s_+ & s_+ & \bar{s}_0\\
s_+^\ast & s_{\uparrow\downarrow} & s_0 & s_-\\
s_+^\ast & s_0^\ast & s_{\uparrow\downarrow} & s_-\\
\bar{s}_0^\ast & s_-^\ast & s_-^\ast & s_\downarrow
\end{array}
\right) \, .
\label{ss}
\end{equation}
These parameters (\ref{ss}) and the conductance parameters (\ref{g})
completely determine the current and noise in the system. Scaling of
all $\hat{r}$ and $\hat{s}$ matrices of a given system by the same
factor does not change the Fano factor.

We proceed with the explicit evaluation of the conductance $\hat{g}$
and shot-noise $\hat{s}$ matrices in the cases of tunnel, diffusive
and ballistic junctions below and then apply our theory to
two-terminal double barrier ferromagnetic-normal-ferromagnetic (\textit{F-N-F})
systems. The results for the conductance matrix $\hat{g}$ have been
derived in Ref.\ \onlinecite{Brataas00:2481} and we briefly reiterate
these results here for completeness. The shot-noise matrix $\hat{s}$
has not been studied before.

\subsubsection{Ballistic junctions}

Because the transverse momentum is conserved in ballistic junctions,
the reflection matrices ($r^{\uparrow}$ and $r^\downarrow$) are
diagonal in the basis of the transverse modes. Simplifying the
situation,\cite{Bauer92:1676} we assume that the diagonal components
of the reflection matrices can attain only two values: Full
transmission $0$ or no transmission $1$. From
Eqs.\ (\ref{LB-1}), (\ref{LB-2}) we see that the Fano factor for a single
ballistic junction vanishes. It is not the case in a double barrier
system, as discussed in the Introduction.

The conductance $\hat{g}$ and shot-noise $\hat{s}$ matrices for a
single junction are in this case 
\begin{eqnarray*}
\hat{g}/g &=&\left(
\begin{array}{cc}
1+p   & 1+|p|\\
1+|p| & 1-p
\end{array}
\right)\,,\\
\hat{s}/g &=&\left(
\begin{array}{cccc}
1+p   & 1+|p| & 1+|p| & 1+|p|\\
1+|p| & 1+|p| & 1+|p| & 1+|p|\\
1+|p| & 1+|p| & 1+|p| & 1+|p|\\
1+|p| & 1+|p| & 1+|p| & 1-p
\end{array}
\right) \, ,
\end{eqnarray*}
where $g=(g^{\uparrow}+g^{\downarrow})/2$ is the average conductance and
$p=(g^{\uparrow}-g^{\downarrow})/(g^{\uparrow}+g^{\downarrow})$ is the
relative polarization. If all the junctions are the same, noise
suppression will depend only on the relative polarization $p$ and the
relative magnetization orientations.

\subsubsection{Diffusive junctions}

The conductance for diffusive junctions can be found with different
methods. One possibility is to solve the diffusion equation with the
proper boundary conditions as was done in
Ref.\ \onlinecite{Brataas00:2481} to find the mixing conductance
$g^{\uparrow\downarrow}$. An alternative approach is to evaluate the
conductance matrix by using random matrix theory (RMT) in the
semiclassical approximation.\cite{Brouwer96:4904,Waintal00:12317} We
will use the latter approach here, because it considerably simplifies
the calculation of the shot-noise parameters.

\begin{figure*}
\includegraphics[scale=0.9]{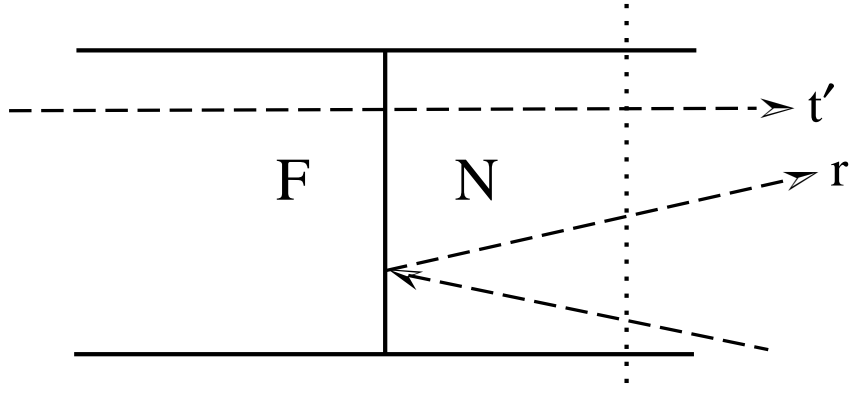}
\caption{A contact between a ferromagnetic node and a normal metal
node. The current is evaluated at the normal metal side (dotted
line). The transmission coefficient from the ferromagnet to the normal
metal is $t^\prime$ and the reflection matrix from the normal metal to
the normal metal is $r$.}
\label{d}
\end{figure*}

We assume that the junction consists of two connected parts as shown
in Fig.\ \ref{d}. The first, normal metal part is described by a
single scattering matrix for both spin-$\uparrow$ and
spin-$\downarrow$ electrons. The second, ferromagnetic part is
described by two independent scattering matrices, one for
spin-$\uparrow$ and one for spin-$\downarrow$ electrons. It is assumed
that there are no correlations between the scattering matrices of the
spin-$\uparrow$ and spin-$\downarrow$ electrons in the ferromagnetic
part. However the up- and down-spin parts of the total scattering
matrix of the combined normal metal and ferromagnetic system are
correlated since both spin directions see the same scattering centers
in the normal metal part. Scattering at the \textit{F-N} boundary is
disregarded since it is assumed that the total resistance is dominated
by the diffuse normal metal and ferromagnetic metal parts of the
junction. The total reflection matrix $r^\alpha$ for spin-$\alpha$
electrons can then be found by concatenating the normal metal and
ferromagnetic parts
\begin{equation}
r^\alpha=r_N+t_N^\prime r_F^\alpha\sum_{n=0}^\infty(r_N^\prime
r_F^\alpha)^nt_N \, .
\label{r}
\end{equation}

We apply the standard polar decomposition for reflection matrices of
the spin-$\alpha$ electrons inside the ferromagnetic
part:\cite{Stone91} $r_F^\alpha=i
U^\alpha\sqrt{T_F^\alpha}U^{\prime \alpha}$, where $U^\alpha$,
$U^{\prime \alpha}$ are $M\times M$ unitary matrices, which in the
isotropic approximation\cite{Stone91} are uniformly distributed in the
group $U(M)$. In the case of time-reversal symmetry, these matrices
are related by transposition, $U^{\prime \alpha}=U^{\alpha^{\rm
T}}$. $T_F^\alpha$ is a diagonal matrix containing the eigenvalues of
$t_F^\alpha (t_F^\alpha)^\dagger$. After inserting these reflection
matrices into the expansion (\ref{r}), we perform the averaging of
traces entering conductance $\hat{g}$ and shot-noise $\hat{s}$
matrices over the scattering matrices in the ferromagnetic part. To
this end we use the semiclassical result for traces of matrices from
RMT.\cite{Brouwer96:4904}

In order to perform the averaging, expressions of the following form
must be evaluated.
\begin{equation}
\rho \equiv \mbox{Tr}[(A\alpha B\beta\cdots\eta
C)(F^\dagger\omega^\dagger\cdots\delta^\dagger E^\dagger\gamma^\dagger
D^\dagger)] \, ,
\label{Tr}
\end{equation}
where $A$, $B$, $C$, $D$, $E$, $F$ are some fixed $M\times M$ matrices
and $\alpha$, $\beta$, $\eta$, $\gamma$, $\delta$, $\omega$ are
unitary matrices uniformly distributed in $U(M)$. It was
shown\cite{Brouwer96:4904} that if the ordered sets {$\alpha$,
$\beta$, $\ldots$, $\eta$} and {$\gamma$, $\delta$, $\ldots$, $\omega$}
are such that any two closest neighbors are independent (or related by
transposition, \textit{i.e.}, $\alpha=\beta^{\rm T}$), then, to the leading
order in $M$, this trace is nonzero only if $A=D$, $B=E$, $\ldots$,
$C=F$ and $\alpha=\gamma$, $\beta=\delta$, $\ldots$, $\eta=\omega$. If
these conditions are satisfied, the trace (\ref{Tr}) equals (to the
leading semiclassical order in the inverse number of transverse modes
$1/M$)
\begin{equation}
\rho =
M\frac{1}{M}\mbox{Tr}[AA^\dagger]\frac{1}{M}\mbox{Tr}[BB^\dagger]\cdots\frac{1}{M}\mbox{Tr}[CC^\dagger]
\, .
\label{RMT}
\end{equation}

This is sufficient to calculate both the conductance $\hat{g}$ and the
shot-noise $\hat{s}$ matrices in terms of the conductance of the
normal metal part and up- and down-spin conductances in the
ferromagnetic part, after we resolve two problems. First, we encounter
the task of calculating the trace $\mbox{Tr}[t_Nt_N^\dagger
t_Nt_N^\dagger]$. Second, the two $\hat{s}$ matrix elements
$\hat{s}_\uparrow=\mbox{Tr}[\hat{1}-r^\uparrow (r^\uparrow)^\dagger
r^\uparrow (r^\uparrow)^\dagger]$ and
$\hat{s}_\downarrow=\mbox{Tr}[\hat{1}-r^\downarrow
(r^\downarrow)^\dagger r^\downarrow (r^\downarrow)^\dagger]$ pose a
challenge, because some of the terms in their expansions [following
Eq.\ (\ref{r})] contain correlations between random matrices which
invalidate the assumptions of the semiclassical result (\ref{RMT}). We
calculate these three quantities using the Boltzmann-Langevin
approach.  It was shown in Ref.\ \onlinecite{Jong95:16867} that in
all-normal metal systems the Fano factor has a universal
$1/3$ suppression in diffusive junctions of arbitrary shape,
dimension, and conductivity distribution. This is directly applicable
to the case at hand because in evaluating $\mbox{Tr}[t_Nt_N^\dagger
t_Nt_N^\dagger]$ and $\mbox{Tr}[r^\alpha (r^\alpha)^\dagger r^\alpha
(r^\alpha)^\dagger]$ we deal with collinear transport. Therefore we
find
\begin{eqnarray}
\mbox{Tr}[t_Nt_N^\dagger t_Nt_N^\dagger]&=&\frac{2}{3}g_N \, ,
\label{sonethird_1} \\ \hat{s}_\uparrow&=&\frac{4}{3}g_{\uparrow} \, ,
\label{sonethird_2} \\ \hat{s}_\downarrow&=&\frac{4}{3}g_{\downarrow} \, .
\label{sonethird_3}
\end{eqnarray}

Now we use the expansion (\ref{r}), the semiclassical result
(\ref{RMT}), and the results (\ref{sonethird_1}), (\ref{sonethird_2}), and
(\ref{sonethird_3}) to find the conductance and shot-noise matrices
\begin{eqnarray}
\hat{g}/g&=&\left(
\begin{array}{cc}
1+p & g_N/g\\
g_N/g & 1-p
\label{csm}
\end{array}
\right),\\ 
\frac{3}{2}\hat{s}/g&=&\left(
\begin{array}{cccc}
2(1+p) & 1+p & 1+p & 0\\ 
1+p & 1 & 1 & 1+p\\ 
1+p & 1 & 1 & 1+p\\ 
0 & 1+p & 1+p & 2(1+p)
\end{array}
\right)\nonumber\\
&&+g_N/g\left(
\begin{array}{cccc}
0 & 1 & 1 & 2\\
1 & 0 & 0 & 1\\
1 & 0 & 0 & 1\\
2 & 1 & 1 & 0
\end{array}
\right)\nonumber\\
&&-g/g_N\left(
\begin{array}{cccc}
  0 & 0 & 0 & 0\\
  0 & 1-p^2 & 1-p^2 & 0\\
  0 & 1-p^2 & 1-p^2 & 0\\
  0 & 0 & 0 & 0
\end{array}
\right) \, ,
\end{eqnarray}
where $g_N$ is the conductance of the normal metal part of the
junction, $g=(g^{\uparrow}+g^{\downarrow})/2$, and
$p=(g^{\uparrow}-g^{\downarrow})/(g^{\uparrow}+g^{\downarrow})$. The
spin-dependent conductance
$g^{\uparrow}=(g_F^{\uparrow}g_N)/(g_F^{\uparrow}+g_N)$ and
$g_F^{\downarrow}=(g_F^{\downarrow}g_N)/(g_F^{\downarrow}+ g_N)$ are
given by Ohm's law in terms of the spin-dependent conductance of the
ferromagnet ($g_F^{\uparrow}$ and $g_F^{\downarrow}$) and the
conductance of the normal metal ($g_N$). Two parameters govern the
Fano factor if all junctions are the same, the relative polarization
$p$ and the fraction of the conductance of the normal metal part to
the average conductance $g_N/g$. The result for the conductance matrix
(\ref{csm}) using RMT agrees with the calculation using the diffusion
equation in Ref.\ \onlinecite{Brataas00:2481}.

\subsubsection{Tunnel junctions}

The transmission coefficients are exponentially small in tunnel
junctions. We expand $\hat{r}$ and $\hat{s}$ in terms of the small
quantities $\delta r^\alpha_{n m}=\delta_{nm}-r^\alpha_{n m}$ and keep only the
lowest order nonvanishing terms. We also assume that the reflection
coefficients have random phases. First-principles band-structure
calculations confirm this for realistic systems.\cite{Xia:unpubl} An
important result can be drawn from the randomness of the phases of
the reflection coefficients: The
imaginary part of the mixing conductance\cite{Xia:unpubl} and, similarly, the shot-noise parameters vanishes.
Using this, we can express the conductance and shot-noise parameters
as
\begin{eqnarray*}
\hat{g}/g &=&\left(
\begin{array}{cc}
1+p & 1\\
1 & 1-p
\end{array}
\right)\,,\\
\hat{s}/g &=&\left(
\begin{array}{cccc}
1+2p & 1+p & 1+p & 1 \\
1+p & 1 & 1 & 1-p\\
1+p & 1 & 1 & 1-p\\
1 & 1-p & 1-p & 1-2p
\end{array}
\right) \, .
\end{eqnarray*}
The parameters only depend on the average conductance $g$ and the
relative polarization of the junctions $p$.

\section{Symmetric \textit{F-N-F} Double Barrier}
\label{s:double}

The theory developed above can be used to analyze the shot noise in
complicated many-terminal devices containing many normal metal nodes
and ferromagnetic reservoirs. Such an analysis has already been
performed for the current through two-terminal double barrier
ferromagnet-normal metal-ferromagnet systems, a novel three-terminal
ferromagnet-normal metal- ferromagnet ``spin-flip'' transistor and
a generalization of Johnson's three-terminal
spin-transistor.\cite{Brataas00:2481} We have seen that, in general,
there are 4 conductance parameters and 11 shot-noise parameters that
completely describe each junction in the system. The Fano factor
defined as the ratio between the shot noise and the current through a
junction thus depends on $15N-1$ parameters, where $N$ is the number
of different junctions. (The Fano factor is invariant under the
scaling of all parameters and, therefore, we can set any one of the
$15N$ parameters to be unity and correspondingly scale the remaining
$15N-1$ parameters.) Clearly, we cannot explore all possible systems,
but we will here illustrate the usefulness of the semiclassical theory
of shot noise to study the most simple system of spin accumulation: A
two-terminal double barrier ferromagnet-normal metal-ferromagnet
system. We will furthermore assume that the system is symmetric and
study the regime of ballistic, diffusive, and tunnel
junctions. Besides, we will investigate the elastic and inelastic
transport regimes separately. The number of independent parameters
(defining the angular dependence of the Fano factor) then reduces to 1
(ballistic), 2 (diffusive), and 1 (tunnel), which considerably
simplifies the analysis. The results for the current and the shot
noise in the collinear configuration, when the magnetizations of the
left and the right probes are parallel or antiparallel, can be easily
deduced from a two channel model of the corresponding all-normal metal
system. All our results in the elastic and inelastic transport regimes
and for ballistic, diffusive, and tunnel junctions agree with these
known results. The results when the magnetizations are noncollinear
are novel. We demonstrate that the Fano factor can exhibit a
nonmonotonic dependence on the relative angle between the
magnetizations in the ferromagnets. This can be used experimentally to
obtain more information about the spin accumulation and the nature of
the junctions including their polarizations.

We assume that both ferromagnetic reservoirs are held at zero
temperature but at different chemical potentials $\mu_0$ and
$\mu_0+|eV|$, where $V$ is the applied constant voltage bias. We
denote $\hat{f}_N^{\text{a}}$ the {\it energy-averaged} distribution
function in the normal metal island for energies between the chemical
potentials of the reservoirs. $\hat{f}_N^{\text{a}}$ is a function of
the conductance matrix $\hat{g}$, the relative angle between the
magnetizations of the ferromagnets, and the applied voltage bias and
can be found by the average charge and spin conservation in the island
as outlined in Ref.\ \onlinecite{Brataas00:2481} and is the same in
both elastic and inelastic regimes.

We use Eq.\ (\ref{c}) for the current flowing through the left and
right junctions and Eqs.\ (\ref{dI}), (\ref{cs}) for the spin and
charge conservation to obtain the form of the fluctuation of the node
distribution function $\delta\hat{f}_N$.  This fluctuation leads to
the total fluctuation in the current given by
\begin{equation}
\Delta I_1(t)=\mbox{Sp}[\hat{V}_1\delta\hat{I}_1(t)+\hat{V}_2\delta\hat{I}_2(t)] \, ,
\label{di}
\end{equation}
with 
$$ 
\hat{V}_n=-\frac{(-1)^n}{2}\hat{1}+\frac{p}{4}\frac{({\bf m}_2-{\bf
m}_1)\eta_r+({\bf m}_2\times{\bf
m}_1)\eta_i}{|\eta|^2\cos^2\frac{\theta}{2}+\eta_r\sin^2\frac{\theta}{2}}
\cdot\hat{\text{\boldmath$\sigma$}} \, ,
$$
where $\eta_r$ ($\eta_i$) is the real (imaginary) part of the relative
mixing conductance $\eta=2g^{\uparrow\downarrow }/(g^{\uparrow
}+g^{\downarrow})$. $\hat{V}_n$ is related to the average zero
temperature distribution function $\hat{f}_N^{\text{a}}$ by $
\hat{V}_n=(\hat{f}_1-\hat{f}_2)(\hat{f}_n-\hat{f}_N^{\text{a}})$,
where $\hat{f}_n$ is the distribution function in the $n$th
ferromagnetic reservoir for energies between the chemical potentials
of the reservoirs. (Because the reservoirs are held at thermal
equilibrium at zero temperature, the allowed values for $\hat{f}_n$
are $\hat{1}$ and $\hat{0}$--the unit $2\times2$ matrix and the zero
$2\times2$ matrix) $\mbox{Sp}$ denotes the trace in spin indices to
distinguish it from the trace in the space of transverse channels
which we denoted $\mbox{Tr}$ in the preceding discussion. Eq.\
(\ref{di}) is valid in both the elastic and the inelastic regimes when
the conductance matrix parameters are energy independent on the scale
defined by the voltage bias.

Using Eq.\ (\ref{di}) we express the time-dependent current
fluctuations in terms of the Langevin sources
$\delta\hat{I}_i(t)$. Then we apply Eq.\ (\ref{cccc}) to find
correlators for these Langevin sources and finally find the shot noise
from Eq.\ (\ref{sn}):
$$
S=\frac{2e^2}{h}\int_{\mu_0}^{\mu_0+|eV|}
dE~\mbox{Sp}\left[\hat{S}_1(E)+\hat{S}_2(E)\right] \, ,
$$
\begin{eqnarray}
\hat{S}_n(E)&=&\sum_{\alpha \beta}\hat{g}^{\alpha
\beta}\{\hat{f}_N(E)[\hat{1}-\hat{f}_N(E)](2\hat{V}_nu_n^{\beta}\hat{V}_nu_n^\alpha\nonumber\\
&&-u_n^{\beta}\hat{V}_n\hat{V}_nu_n^\alpha)-[\hat{f}_N(E)-\hat{f}_n]^2u_n^{\beta}\hat{V}_n\hat{V}_nu_n^\alpha\}\nonumber\\
&&+\sum_{\alpha \alpha^\prime \beta
\beta^\prime}\hat{s}^{\alpha\beta\alpha^\prime\beta^\prime}\{u_n^\alpha[\hat{f}_N(E)-\hat{f}_n]\nonumber\\
&&\times u_n^{\alpha^\prime}\hat{V}_nu_n^\beta[\hat{f}_N(E)-\hat{f}_n]u_n^{\beta^\prime}\hat{V}_n\} \, .
\label{s}
\end{eqnarray}
In each sum $\alpha$, $\alpha^\prime$, $\beta$, and $\beta^\prime$ are
spin indices, $u_n^\alpha=(\hat{1}+\alpha {\bf m}_n\cdot
\hat{\text{\boldmath $\sigma$}})/2$ are projection matrices
corresponding to the magnetization direction ${\bf m}_n$ of the
$n$th reservoir, $\hat{f}_N(E)$ is the distribution function
on the normal metal island and
$\hat{g}^{\alpha\beta}=\mbox{Tr}[\hat{1}-r^{\alpha}(r^\beta)^\dagger]$
and
$\hat{s}^{\alpha\beta\alpha^\prime\beta^\prime}=\mbox{Tr}[\hat{1}-r^{\alpha}(r^{\alpha^\prime})^\dagger r^{\beta}(r^{\beta^\prime})^\dagger]
$ are elements of the conductance matrix $\hat{g}$ (\ref{g}) and shot
noise $\hat{s}$ matrix (\ref{spara}).

All the terms in the sums of Eq.\ (\ref{s}) are not independent: The
elements of the first sum can be grouped to form a $2\times2$
Hermitian matrix while the elements of the second sum can be grouped
to form a $4\times4$ Hermitian matrix with the same structure as the
shot-noise $\hat{s}$ matrix (\ref{ss}).

In the following subsections we directly apply Eq.\ (\ref{s}) to the
cases of ballistic, diffusive, and tunnel junctions in both elastic
and inelastic regimes.

\subsection{Elastic transport}

\subsubsection{Ballistic junctions}

\begin{figure*}
\includegraphics[scale=0.9]{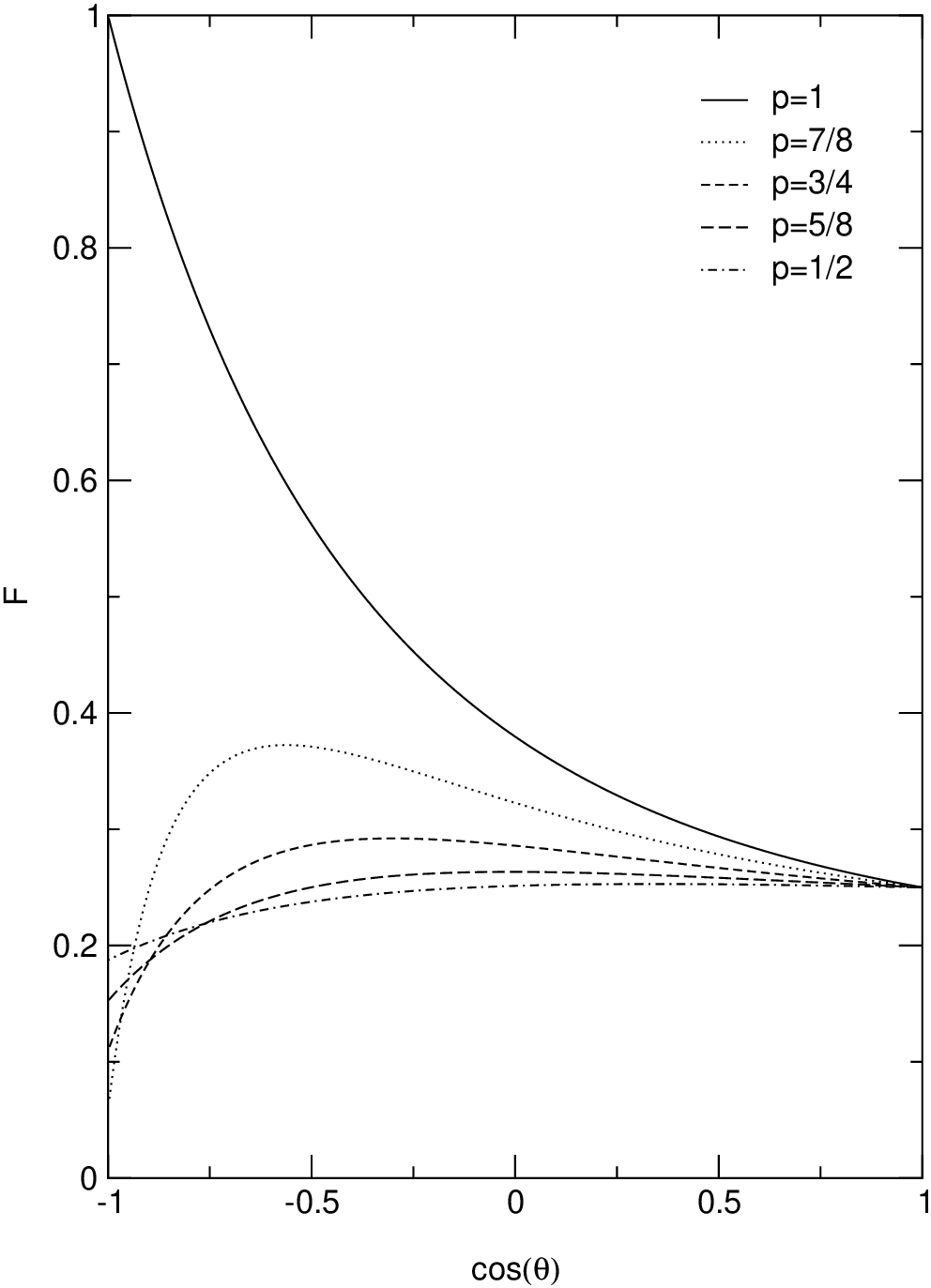}
\caption{Angular dependence of the Fano factor $F$ in \textit{F-N-F} systems in
the elastic regime with ballistic junctions. The results are shown for
different polarizations
$p=\left(g^{\uparrow}-g^{\downarrow}\right)/\left(g^{\uparrow}+g^{\downarrow}\right)$.}
\label{e-b}
\end{figure*}

The Fano factor for the collinear configurations of the ferromagnetic
magnetizations can be found by applying the result for all-normal
metal systems (\ref{F_ET2}) for two independent spin channels
\begin{eqnarray*}
F(\theta=0)&=&\frac{1}{4} \, ,\\
F(\theta=\pi)&=&\frac{1-p^2}{4} \, .
\end{eqnarray*}
The Fano factor is decreased in the antiparallel configuration
$\theta=\pi$ when each spin component experiences an asymmetric double
barrier system. The angular dependence of the Fano factor is shown in
Fig.\ \ref{e-b} for a set of different polarizations $p$.

For polarizations $p$ below the critical value
$p_c=1/3$, the angular dependence of the Fano factor
is monotonic. When $p$ exceeds $p_c$, there is a maximum in $F$ which
continuously (and monotonically) increases from $\theta=0$ to
$\theta=\pi$ as $p$ increases from $p_c$ to $1$. $F_{\rm max}(p)$
monotonically increases from $1/4$ to $1$ when $p$ is increased from
$p_c$ to $1$. The position of the maximum thus gives another
independent measurement of the polarization $p$ together with the
ratio $F(\theta=\pi)/F(\theta=0)$.

It is interesting to note the limiting behavior of $F$ when $\theta$
approaches $\pi$ and $p$ approaches $1$: $\lim_{p\rightarrow
1}\lim_{\theta\rightarrow\pi}F(\theta,p)=0$ and
$\lim_{\theta\rightarrow\pi}\lim_{p\rightarrow 1}F(\theta,p)=1$. Thus,
for half-metallic ferromagnets ($p=1$), there is a sharp drop in $F$
from $1$ to $0$ when $\theta$ approaches $\pi$.

In the exact antiparallel configuration, $\theta=\pi$, the Fano factor
vanishes when the system is close to being half-metallic ($p$ is close
$1$) since the transport properties are governed by the barrier with
the lower transmission which has a vanishingly small Fano factor for
ballistic junctions. $p\approx1$ means that one barrier has vanishing
conductance and, therefore, the current and the shot noise vanish in
the system. Relaxing the assumption of antiparallel magnetization when
the relative angle is slightly below $\pi$, the Fano factor increases
to $1$ as some of the low transmission channels form. These low
transmission channels determine the transport properties and the
system effectively behaves as a single tunnel barrier system with
$F\approx1$.

\subsubsection{Diffusive junctions}

\begin{figure*}
\includegraphics[scale=0.9]{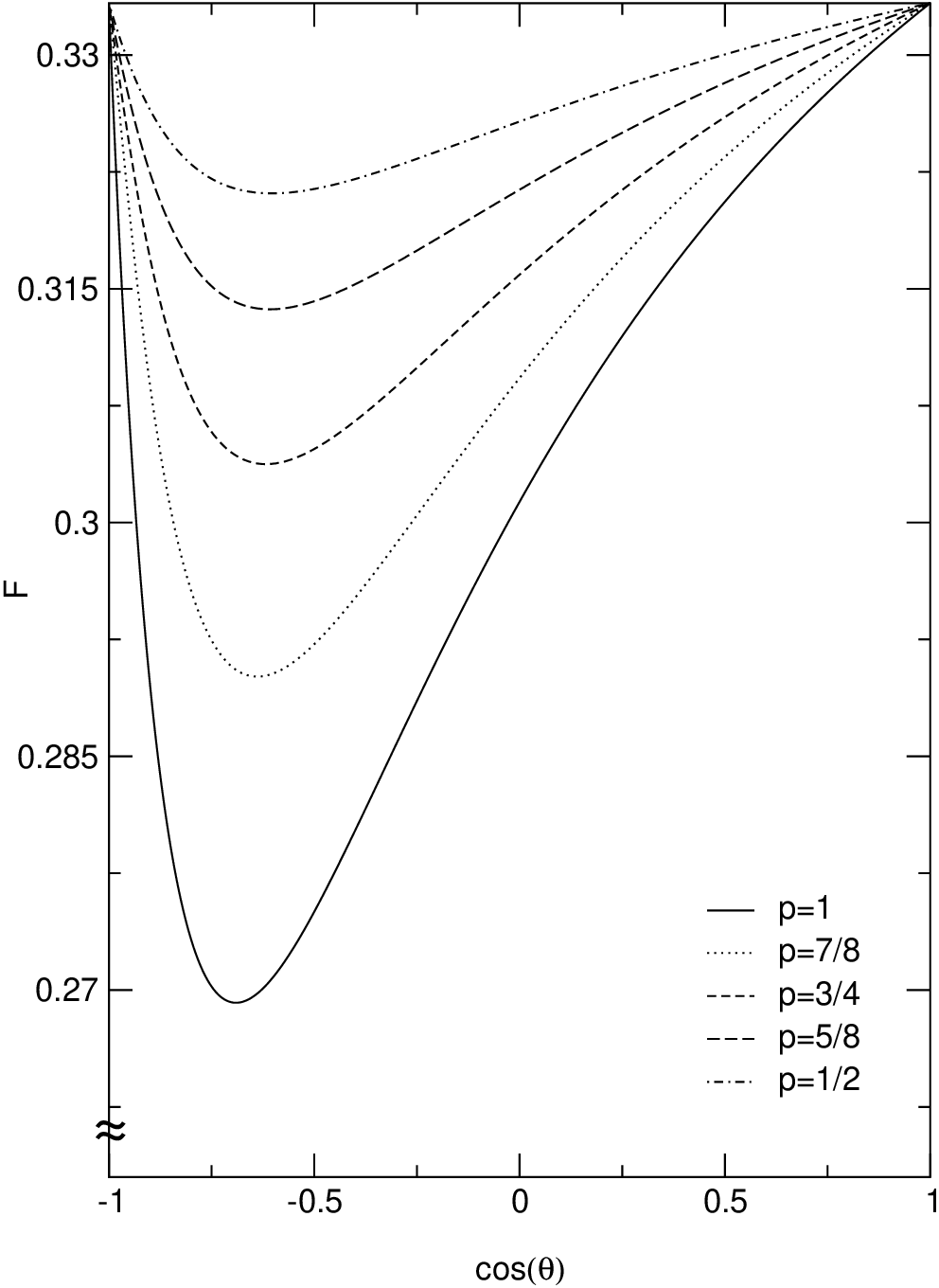}
\caption{Angular dependence of the Fano factor $F$ in \textit{F-N-F} systems in
the elastic regime with diffusive junctions. The results are shown for
different polarizations
$p=\left(g^{\uparrow}-g^{\downarrow}\right)/\left(g^{\uparrow}+g^{\downarrow}\right)$. $g_N/g=4$.}
\label{e-d}
\end{figure*}

As discussed in the Introduction, the Fano factor has the universal
value of $1/3$ in the collinear configuration for diffusive
junctions. Thus we have from Eq.\ (\ref{F_ED2}) for two spin-channels
\begin{eqnarray*}
F(\theta=0)   & = & \frac{1}{3} \, ,\\
F(\theta=\pi) & = & \frac{1}{3} \, .
\end{eqnarray*}
The $1/3$ suppression also holds for arbitrary angles when $p=0$ so
that there is no spin accumulation. In general, there is an angular
dependence of the Fano factor since the spins are coherent on the
normal metal portion of the junctions. We show this dependence in
Fig.\ \ref{e-d} for $g_N/g=4$ and several values of $p$. When $p$ is
finite, there is a minimum in $F(\theta)$. The position of the minimum
is relatively insensitive to the value of the polarization $p$. When
$g_N/g$ increases, this minimum shifts to larger angles and the Fano
factor flattens in the limit of $g_N/g\rightarrow\infty$ (in this
limit the junctions are fully ferromagnetic): $F(\theta)=1/3$ for
$\theta<\pi$ with a singularity at $\theta=\pi$ when $F$ has a sharp
minimum which deepens to zero for $p=1$. This mimics the behavior of
the ballistic systems. The Fano factor identically equals $1/3$ in a
special case of $p=1$ and $g_N/g=2$ when
$g_F^\uparrow/g_N\rightarrow\infty$ and $g_F^\downarrow/g_N\rightarrow
0$, where $g_F^\alpha$ is spin-$\alpha$ conductance of the
ferromagnetic portion of the junction [for a diffusive junction, in
general, $g^\alpha=g_F^\alpha g_N/(g_F^\alpha+g_N)$].

\subsubsection{Tunnel junctions}

\begin{figure*}
\includegraphics[scale=0.9]{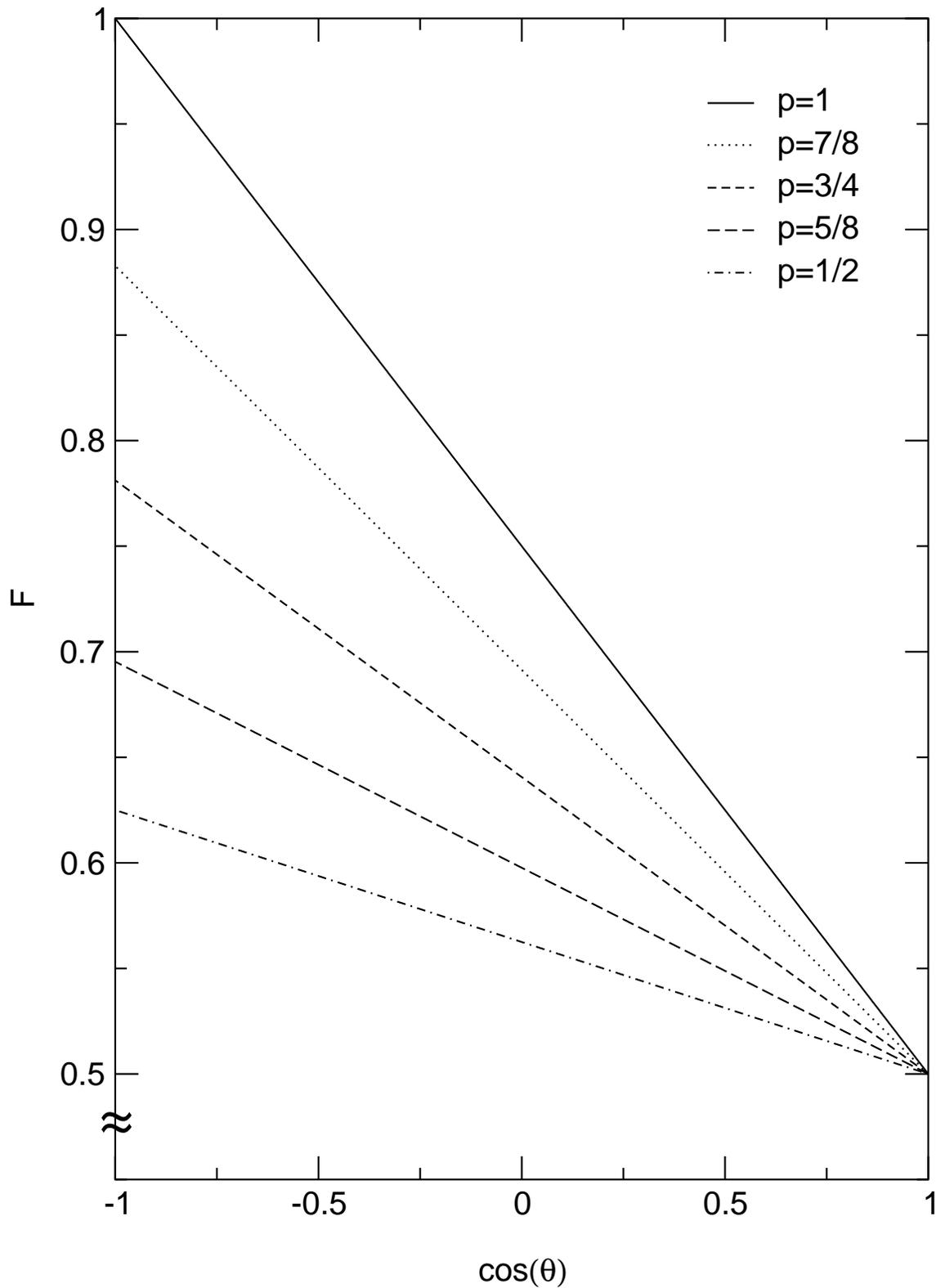}
\caption{Angular dependence of the Fano factor $F$ in \textit{F-N-F} systems
with tunnel junctions. The results are shown for different
polarizations
$p=\left(g^{\uparrow}-g^{\downarrow}\right)/\left(g^{\uparrow}+g^{\downarrow}\right)$.}
\label{i-t}
\end{figure*}

In this regime, the Fano factor displays a simple analytic dependence
on the angle $\theta$ for any polarization $p$:
\begin{equation}
F(\theta)=\frac{1}{2}\left(1+p^2\sin^2\frac{\theta}{2}\right) \, .
\label{Ft}
\end{equation}
The angular dependence is thus a simple monotonic interpolation
between the values of the Fano factor in the collinear configurations
[from Eq.\ (\ref{F_EB2}) for two spin-channels]
\begin{eqnarray*}
F(\theta=0)&=&\frac{1}{2} \, ,\\
F(\theta=\pi)&=&\frac{1+p^2}{2} \, .
\end{eqnarray*}
For completeness, we plot the corresponding Fano factor in Fig.\
\ref{i-t}.

\subsection{Inelastic transport}

\subsubsection{Ballistic junctions}

The collinear configuration in this regime gives a vanishing Fano
factor
\begin{eqnarray*}
F(\theta=0)   & = & 0 \, ,\\
F(\theta=\pi) & = & 0 \, .
\end{eqnarray*}
Also when the polarization $p=0$, the Fano factor $F$
vanishes for any angle $\theta$. But in the case of noncollinear
configuration and nonzero polarization $p$, the Fano factor $F$ is
always finite. This can be understood as follows: In general, the
time-dependent distribution function of the normal metal node has a FD
form for the two spin orientations along some (time-dependent)
direction ${\bf u}(t)$. When there is a spin accumulation on the node,
the chemical potentials corresponding to the electrons with spin
parallel and antiparallel to ${\bf u}(t)$ are different. The
electrons are injected into the normal metal island with spin
orientations along the magnetization direction of the corresponding
reservoir, ${\bf m}_1$ or ${\bf m}_2$. This means that these projected
components of the distributions functions may be partially
occupied. Partial occupation of the states effectively results in a
'thermal noise' contribution to the shot noise of the
system. Consequently, the Fano factor is finite for intermediate
angles $\theta$. This mimics the mechanism for nonzero shot noise in
the all-normal metal elastic double barrier system with ballistic
junctions which we discussed in the Introduction.

\begin{figure*}
\includegraphics[scale=0.9]{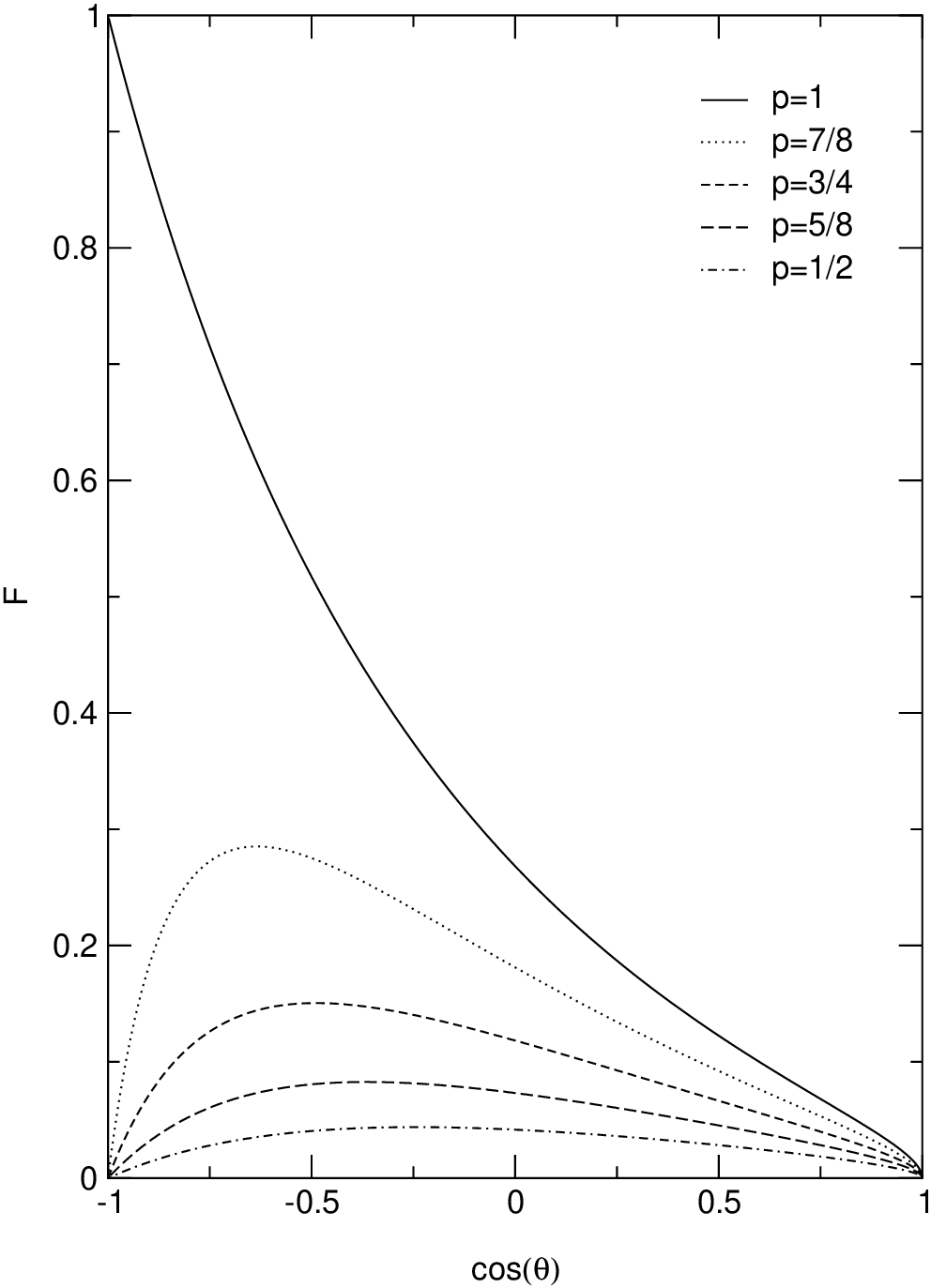}
\caption{Angular dependence of the Fano factor $F$ in \textit{F-N-F} systems in
the inelastic regime with ballistic junctions. The results are shown
for different polarizations
$p=\left(g^{\uparrow}-g^{\downarrow}\right)/\left(g^{\uparrow}+g^{\downarrow}\right)$.}
\label{i-b}
\end{figure*}

The analytic form of the Fano factor $F(\theta)$ is relatively
simple:
$$
F(\theta) =
\frac{p^2\sin\theta\cos\frac{\theta}{2}\left[1+|p|\left(1+2\sin\frac{\theta}{2}\right)\right]}{8\left(1-|p|\sin^2\frac{\theta}{2}\right)\left(1+|p|\cos^2\frac{\theta}{2}\right)^2}
\, .
$$
In Fig.\ \ref{i-b} we show the angular dependence of the Fano factor
for a set of different polarizations $p$. When $p$ increases from $0$
to $1$, $F(\theta)$ yields a single maximum which monotonically shifts
from $\theta=0$ to $\theta=\pi$. $F_{\rm max}(p)$ monotonically
increases from $0$ to $1$, when p is varied from $0$ to $1$. The
limiting behavior found in the elastic regime is also valid in the
inelastic regime.

\subsubsection{Diffusive junctions}

\begin{figure*}
\includegraphics[scale=0.9]{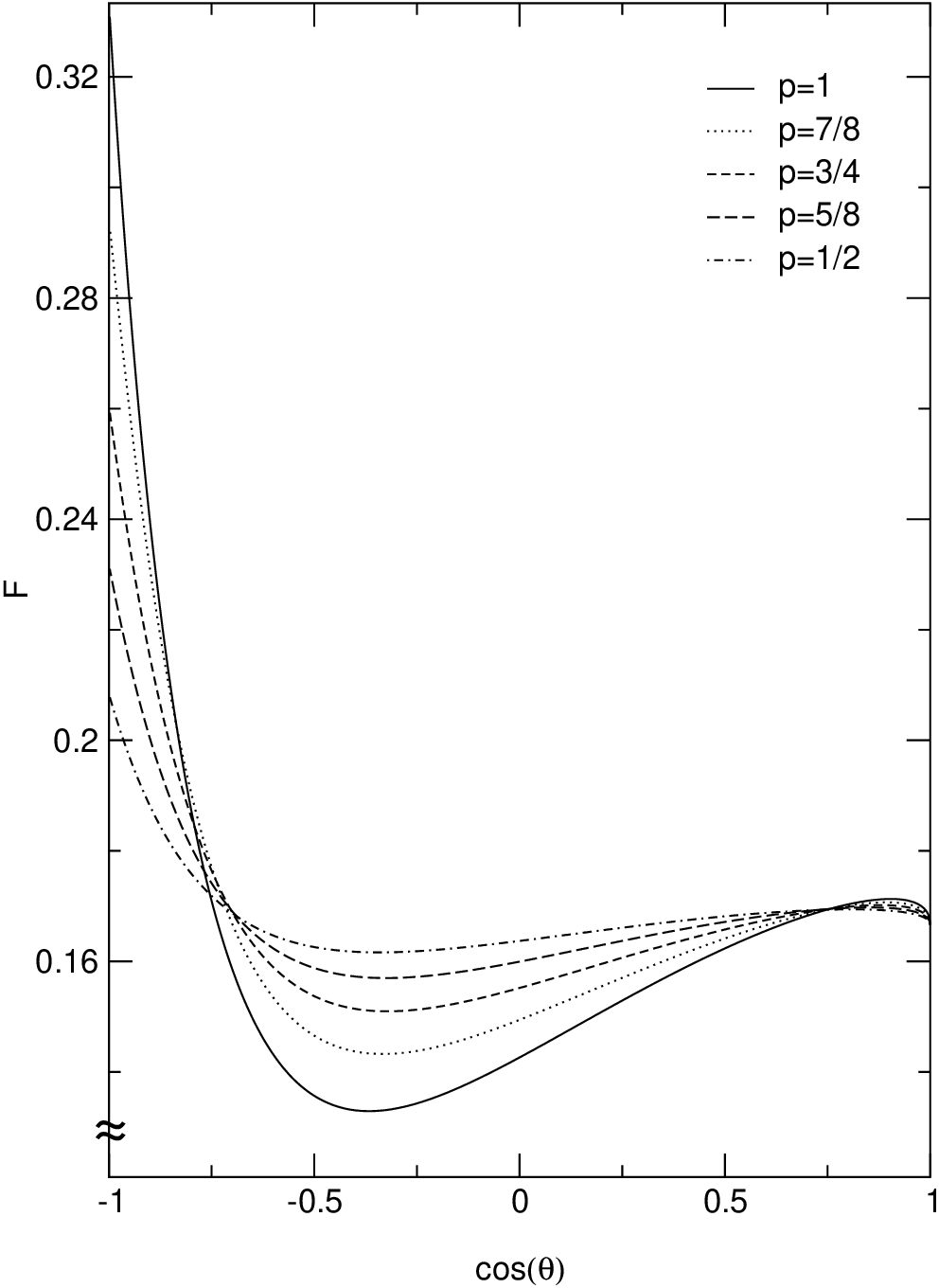}
\caption{Angular dependence of the Fano factor $F$ in \textit{F-N-F} systems in
the inelastic regime with diffusive junctions. The results are shown
for different polarizations
$p=\left(g^{\uparrow}-g^{\downarrow}\right)/\left(g^{\uparrow}+g^{\downarrow}\right)$. $g_N/g=4$.}
\label{i-d}
\end{figure*}

In the collinear configuration the Fano factor is
\begin{eqnarray*}
F(\theta=0)&=&\frac{1}{6} \, ,\\
F(\theta=\pi)&=&\frac{1+p^2}{6} \, .
\end{eqnarray*}
The $1/6$ suppression also holds for any $\theta$, when $p=0$. We show
the angular dependence of the Fano factor for a set of different
polarizations $p$ and $g_N/g=4$ on Fig.\ \ref{i-d}. For nonzero $p$,
$F(\theta)$ has two local extrema, one minimum and one maximum, the
minimum corresponding to a higher $\theta$. As in the elastic case,
the angles corresponding to the extrema in $F(\theta)$ are relatively
insensitive to the value of the polarization $p$.  The function
$F(\theta)$ becomes monotonic for sufficiently small $g_N/g$.  In the
opposite limit, when this ratio goes to infinity,
$g_N/g\rightarrow\infty$, the junctions become completely
ferromagnetic. In this limit, $F(\theta)=1/6$ for $\theta<\pi$ with a
jump to $(1+p^2)/6$ at $\theta=\pi$ much like in the elastic regime:
Before the steep rise, the Fano factor $F$ has a sharp minimum at
$\theta=\pi$ which deepens to zero for $p=1$.

\subsubsection{Tunnel junctions}

As was noted in the Introduction for the case of all-normal metal
systems, the elastic and inelastic transport regimes yield the same
Fano factor. We obtain that this remains valid for \textit{F-N-F} systems even
in the case of noncollinear configuration, \textit{i.e.} $F$ is given by
Eq.\ (\ref{Ft}).

\section{Conclusions and Discussions}
\label{s:con}

A semiclassical circuit theory for the shot noise in multiterminal
ferromagnet-normal metal systems has been developed. The current is
governed by 4 real parameters for each contact and the shot noise is
governed by 11 real parameters. The shot-noise parameters have been
evaluated for diffusive, tunnel, and ballistic junctions.

The circuit theory has been applied to symmetric double barrier \textit{F-N-F}
systems.  Both the elastic and inelastic regimes have been studied. In
each of the regimes, a complex nonmonotonic angular dependence of the
Fano factor was found in the case of diffusive junctions. In the case
of ballistic junctions with large conductance polarization, a sharp
drop of the Fano factor from 1 to 0 was found when the magnetization
of the ferromagnets approaches the antiparallel configuration. Simple
sinusoidal angular dependence of the Fano factor was found for the
tunnel junctions in both the elastic and inelastic regimes.

We have in this work disregarded spin-flip processes. Spin-flip
scattering can occur both in the junctions and on the nodes and these
different processes will influence the current and the shot noise in
different ways. The spin-flip processes on the nodes can be
disregarded when the spin-flip relaxation time is longer than the
transport dwell time $\tau_{\text{sf}} \gg \tau_d$. For shorter
spin-flip scattering times, the angular dependence of the average
current is governed by a reduced effective polarization and a reduced
mixing conductance.\cite{Brataas00:2481} We expect similarly that the
shot noise is governed by a reduced effective polarization and a
reduced effective mixing conductance.  The anticipated effect of
spin-flip scattering in the normal metal node on the Fano factor is
thus mostly quantitative; the effective polarization and mixing
conductance are reduced making the angular dependence weaker, and the
angular dependence eventually vanishes when $\tau_{\text{sf}} \ll
\tau_d$. (A discussion of the spin-flip scattering time
$\tau_{\text{sf}}$ \textit{vs} the transport dwell time $\tau_d$ has been
given by Brataas \textit{et al.} in Ref.\ \onlinecite{Barnas98:85}). Spin-flip
scattering in the junctions can be disregarded when the junction is
smaller in the transport direction than the spin-flip scattering
length. If that is not the case, spin-flip scattering in the junction
can have a qualitative effect as well as a quantitative effect both on
the current and the shot noise. First, we also expect that the
effective polarization and mixing conductance decrease with increasing
spin-flip scattering in the contact. There is also a second,
qualitative effect: In the absence of spin-flip scattering in the
junctions, the current between a ferromagnetic node and a normal metal
is determined by a $2 \times 2$ conductance matrix (\ref{g}) and the
fluctuations in the current are governed by a $4 \times 4$ shot-noise
matrix (\ref{spara}). However, these matrices do not completely
determine the current and shot noise in the presence of spin-flip
scattering in the junction since more combinations of the scattering
matrices become accessible (for the calculation of the average current
see e.g. the appendix in Ref.\
\onlinecite{Brataas00:2481}). Consequently, an expanded basis of
conductance and shot-noise parameters must be used and the angular
dependence of both the current and the shot noise can be qualitatively
different than in the case of no spin-flip scattering in the
junctions.

\acknowledgments

We are grateful to G.\ E.\ W.\ Bauer, W.\ Belzig, B.\ I.\ Halperin, D. Huertas Hernando and Yu.\ V.\ Nazarov for stimulating discussions. This work was supported in part by the Norwegian Research Council, NSF Grant No. DMR 99-81283, the Schlumberger Foundation, and the NEDO International Joint Research Grant Program ``Nano-magnetoelectronics'',

\appendix

\end{document}